\documentclass[12pt]{article}
\usepackage[nohyper]{alpha}
\usepackage{amsmath,amsfonts,epsfig,cite,graphics}
\usepackage{amssymb}
\usepackage{mathrsfs} 
\usepackage{amsbsy,color}
\usepackage{latexsym}
\usepackage{graphicx}
\usepackage{psfrag}
\usepackage{multirow}
\usepackage{booktabs}
\usepackage{url}
\usepackage{dcolumn}
\usepackage[T1]{fontenc}

\usepackage{soul}

\newcommand{\be}{\begin{equation}}
\newcommand{\ee}{\end{equation}}
\newcommand{\bea}{\begin{eqnarray}}
\newcommand{\eea}{\end{eqnarray}}
\newcommand{\bes}{\begin{eqnarray}}
\newcommand{\ees}{\end{eqnarray}}
\newcommand{\ba}{\begin{array}}
\newcommand{\ea}{\end{array}}

\newcommand{\pref}[1]{(\ref{#1})}

\newcommand{\Eq}[1]{Eq.~(\ref{#1})}

\newcommand{\Fig}[1]{Figure~\ref{#1}}

\newcommand{\Sect}[1]{Section~\ref{#1}}

\newcommand{\Tab}[1]{Table~\ref{#1}}

\newcommand{\Ref}[1]{Ref.~\cite{#1}}
\newcommand{\Refs}[1]{Refs.~\cite{#1}}

\newcommand{\nneg}{n_\mathrm{neg}}

\newcommand{\csw}{c_\mathrm{\tiny SW}}

\newcommand{\fm}{\mathrm{fm}}

\pdfoutput=1   

\begin{document}

\preprintno{
DESY 20-041\\
MITP/20-010
}

\title{Remarks on strange-quark simulations with Wilson fermions}

\author[him,gsi,jgu]{Daniel~Mohler}
\author[desy]{Stefan~Schaefer}
%
%
\address[him]{Helmholtz-Institut Mainz, Johannes Gutenberg-Universit\"at,
D-55099 Mainz, Germany}
\address[gsi]{GSI Helmholtzzentrum f\"ur Schwerionenforschung, Darmstadt, Germany}
\address[jgu]{Johannes Gutenberg-Universit\"at Mainz, D-55099 Mainz, Germany}
\address[desy]{John von Neumann Institute for Computing (NIC), DESY \\
    Platanenallee 6, D-15738 Zeuthen, Germany}
\begin{abstract}
  In the simulation of QCD with 2+1 flavors of Wilson fermions, the positivity of the fermion
  determinant is generally assumed. We present evidence that this assumption is in general not
  justified and discuss the consequences of this finding.
\end{abstract}


\maketitle

\section{Introduction}
A widespread choice for lattice QCD simulations is a setup with two light
mass-degenerate quarks to which a strange quark and possibly a charm quark are added.
The latter two are typically included with algorithms like the RHMC~\cite{Clark:2006fx} or 
the PHMC~\cite{Frezzotti:1998eu,Frezzotti:1998yp}, which effectively take
the modulus of the fermion determinant of the strange and the charm quark.
In QCD simulations, it is typically assumed that the masses of the
  strange and the charm are large enough, and that therefore the use of the
  modulus has no effect.

In general, chiral symmetry together with $\gamma_5$-Hermiticity
guarantees the positivity of the fermion determinant for each flavor, but for
Wilson fermions the explicit breaking of chiral symmetry makes negative values
possible. Thus, in the absence of further restrictions, there are regions of
configuration space, where the fermion determinant is negative. The
assumption of its positivity is based on the
belief that such regions are not part of those drawn by importance sampling
algorithms in an ensemble with typical statistics.
For sure, one would expect that towards the continuum limit the probability of
such configurations decreases rapidly.

In this paper we discuss the observation of negative fermion determinants in
large scale simulations undertaken by the Coordinated Lattice Simulation (CLS)
effort~\cite{Bruno:2014jqa,Bruno:2016plf}. The CLS consortium has generated ensembles
with 2+1 flavors of non-perturbatively improved Wilson fermions with 
lattice spacings ranging from $0.09$~fm to $0.04$~fm and quark masses including
the physical light- and strange-quark masses.
One particularity is that most simulation points are along lines of constant 
sum of bare quark masses, tuned such that this sum equals the sum
of the physical quark masses. This means that at the symmetric point, 
the quark masses are roughly a third of the physical strange. Only 
at physical light-quark masses, also the third quark attains the mass
of the physical strange.

These simulations employ the {\tt openQCD} code~\cite{openQCD} whose general
algorithmic setup, including twisted-mass reweighting~\cite{Luscher:2008tw}
for the light quarks and the RHMC algorithm for the strange~\cite{Clark:2006fx}, is
described in \Ref{Luscher:2012av}. Since these algorithmic choices, in
particular the modification to the action used by the twisted-mass reweighting
and in the RHMC, could have an impact on the observed spectra of the Dirac
operator, we summarize them in the following, also giving details which so far
have not been published.

To our knowledge, there has not been a detailed analysis of this kind of problem
in the literature, which might also be related to the fact that among the major 
discretizations, only Wilson-type fermions are affected. Note that in the non-QCD lattice literature,
such problems have previously been discussed, see
e.g.~\cite{Bergner:2011zp}. In \Sect{s:light} we briefly investigate the potential problem of 
autocorrelations due to the (twisted-mass regularized) light quark
determinant, but do not find an additional problem on the two ensembles
analyzed.

\section{Description of the problem}
The partition sum for $2+1$ flavor simulations  is given by the 
path integral over the gauge field variables $U$ 
\begin{equation}
Z=\int
  [dU]  \det\{D(m_\mathrm{ud})\}^2 \  \det\{D(m_\mathrm{s})\} \,\mathrm{e}^{-S_\mathrm{g}[U]} \,.
\end{equation} 
Here $D(m)$ is the (improved) Wilson Dirac operator~\cite{Wilson,impr:SW}, whose gauge field
dependence is implicit,
\begin{equation} 
D(m_0) = 
\frac{1}{2}\sum_{\mu=0}^3 \{\gamma_\mu(\nabla^*_\mu+\nabla_\mu) -a\nabla^*_\mu\nabla_\mu\}
+a\csw \sum_{\mu,\nu=0}^3\frac{i}{4}\sigma_{\mu\nu}\widehat F_{\mu\nu}+m_0\,.
\end{equation}
 It depends on the quark mass, $m_\mathrm{ud}$ for the light and $m_\mathrm{s}$ 
for the strange,
with $\nabla_\mu$ and $\nabla_\mu^*$ the covariant
forward and backward derivatives, respectively. The improvement term
containing the standard discretization of the field strength
tensor $\widehat F_{\mu\nu}$~\cite{impr:pap1} comes with the coefficient $\csw$.
For the CLS setup with a tree-level improved gauge action $S_\mathrm{g}$,
the coefficient $\csw$ has been determined in \Ref{Bulava:2013cta}.

Because of the $\gamma_5$-Hermiticity of the Dirac operator $\gamma_5 D \gamma_5 =
D^\dagger$, the determinant is real, $\det\{D\}=(\det\{D\})^*$, and each eigenvalue
$\lambda$ of the Dirac operator 
 is either real or comes in a complex conjugate pair, i.e.  $\lambda^*$ is also
an eigenvalue of $D$.
Since the determinant is the product of all eigenvalues,
negative values of the fermion determinant  appear if there is an odd
number $n_\mathrm{neg}$ of negative real eigenvalues of $D$.

For practical simulations there are  two issues which arise from the
observation of negative, real eigenvalues of the strange Dirac operator. On the
one hand, one needs to be aware of the fact that the determinants themselves are
not suitable weights for the Monte Carlo evaluation of the path integral,
because they are not positive. On the other hand, important regions in field
space are connected by regions with small weight, which is a challenge for
typical update algorithms. These two issues are discussed now.

Before we go into details, we note that it is standard to use an even-odd
decomposition of the fermion determinant~\cite{DeGrand:1988vx}. Deriving from a two-color labeling of the lattice
sites, we have
\be
\det \{D\} = \det \{D_\mathrm{oo}\} \det \{D_\mathrm{ee} - D_\mathrm{eo} D_\mathrm{oo}^{-1} D_\mathrm{oe}\} \equiv \det \{D_\mathrm{oo}\} \det \{\hat D\}
\label{e:eo}
\ee
if $D_\mathrm{oo}(x)$ is invertible for all sites $x$. In our simulations it is checked that  $\det \{D_\mathrm{oo}(x)\}>0$ on
all configurations. Since $\hat D$ is $\gamma_5$-Hermitian, the
above discussion also applies  to this operator and the negativity of the  fermion determinant can be equally analyzed with $\hat D$.
The number of negative eigenvalues of $D$ and $\hat D$ is the same.

\subsection{Monte Carlo}
Virtually all large scale lattice simulations use a Markov Chain Monte Carlo,
where the probability with which a configuration is drawn is given by the 
action terms 
\be
P(U) \propto \det \{D_\mathrm{ud}\}^2 |\det \{D_\mathrm{s}\}| e^{-S_g[U]}\equiv e^{-S_g[U]-S_{f,ud}[U]-S_{f,s}[U]}\, .
\label{eq:w1}
\ee
If the determinant of $D_\mathrm{s}$ is not manifestly positive,
one therefore needs to include a reweighting factor
\be
W_s=\frac{\det \{D_\mathrm{s}\}}{|\det \{D_\mathrm{s}\}|}=(-1)^{n_\mathrm{neg}}
\ee
in the measurement.
Expectation values can then be computed in the standard way using~\cite{PhysRevLett.61.2635}
\be
\langle A \rangle = \frac{\langle A W_s \rangle_+}{\langle W_s \rangle_+}\,,
\ee
with $\langle \cdot \rangle_+$ the expectation value in the theory with the modulus of
the strange determinant taken, as in \Eq{eq:w1}.

The determination of $n_\mathrm{neg}$ is the subject of \Sect{sec:neg}
and is numerically expensive. The inclusion of such a fluctuating sign has an
impact on the uncertainties which can be reached in the actual
simulation. The impact depends strongly on the covariance between the observable and the reweighting factor.

\subsection{Update algorithms}
The second problem is the performance of the typical update algorithms which move in small steps 
in configuration space, like the molecular dynamics based Hybrid Monte Carlo~\cite{Duane:1987de}.
Since they change the gauge fields quasi-continuously, changing $n_\mathrm{neg}$ by one unit can
only be achieved by going through configurations with a zero eigenvalue of $D$. A change by two 
units is possible if a pair of complex conjugate eigenvalues with negative real parts approach
the $\mathrm{Im}(\lambda)=0$ axis and become real. 

From a vanishing real eigenvalue of $D(m)$ follows a vanishing fermion determinant, which means
that one needs to pass through configurations $U$ with  vanishing weight $P(U)$ as 
given in \pref{eq:w1}. Configuration space is therefore divided into
sectors of even and odd $n_\mathrm{neg}$, where $\nneg$ is in general different for light and
strange quarks.

For exact integration of the molecular dynamics equations of motion, which are at the base of the 
HMC algorithm, this means that changing $n_\mathrm{neg}$ by one unit is highly unlikely if not impossible.
While in a practical simulation this integration is never exact, it will at least lead to very
long autocorrelation times, with long periods in which $n_\mathrm{neg}$ is constant.

Since non-vanishing $\nneg$ is a discretization effect, most of the time
$\nneg$ will be zero. Even in a situation, where no nonvanishing values have
been observed on a given Markov-chain, the question arises if this is a
consequence of poor sampling, or if indeed the probability of non-zero values
is so small that this is a likely outcome.

\subsection{Twisted mass reweighting}

For the light quarks, the problem discussed in the previous section has been already identified
and addressed in \Ref{Luscher:2008tw}. The proposed method has also been used in
the CLS simulations.
Instead of generating gauge field configurations with the contribution of the two light fermions
to the weight  given by $\det \{D^\dagger(m)D(m)\}$, one uses even-odd preconditioning \ref{e:eo}
and for the determinant of the Schur complement
\be
\det\{\hat D^\dagger(m)\hat D(m)\} \to \det\left\{\frac{[ \hat D^\dagger(m)\hat D(m)+\mu_0^2 ]^2 }{\hat D^\dagger(m)\hat D(m)+2\mu_0^2}\right\}\,.
\label{eq:tm}
\ee
The ratio between this weight function and $\det \{\hat D^\dagger(m)\hat D(m)\}$ as it appears in the path 
integral is then included by a reweighting factor into the measurement.

The twisted mass parameter needs to be chosen with care. On the one hand, it  should be large enough
to lower the barrier in the action and make all of configuration space accessible also in practical terms.
On the other hand, it has to be small enough such that the fluctuations of the reweighting factor do not
induce too much noise in the measurement and statistical uncertainties are kept under control.

\subsection{RHMC}
In the CLS simulations, also the strange quark is included with an approximation to the corresponding
fermion determinant
\be
\det \{\hat D\}   \to \frac{1}{\det \{R(\hat D^\dagger \hat D)\}}\,,
\ee
where $R(x)$ is a rational approximation to the inverse square root. The product between the two determinants
needs to again be included, by reweighting in the measurement. For its computation one assumes that 
the factor is positive at the strange-quark mass; an assumption which turns out to be incorrect for our ensembles.

The Zolotarev rational function has three parameters: the upper and lower bound of the approximation as well
as the number of poles used. Between the bounds the function then has a defined maximal error.
In general, one aims at a situation where the reweighting factors fluctuate little and 
there are no eigenvalues of the matrix outside of the bounds.

We note, however, that the rational function also provides a cut-off for the
action $S_\mathrm{s,f}=\mathrm{tr} \log  R$, which even for vanishing
eigenvalues of $D$ stays finite. This
  approach can therefore also avoid a sector formation and possible practical problems of ergodicity.

\section{CLS simulations}

Within the CLS effort, a sizeable library of gauge field configurations with
2+1 flavors of improved Wilson fermions has been generated using the algorithmic
setup discussed in the previous section and \Refs{Luscher:2011kk,Bruno:2014jqa}. The lattice spacing ranges from $a\approx 0.09$~fm
down to $a\approx 0.04$~fm, with pion masses down to the physical masses, see
\Tab{tab:ens} for an overview of a subset of these ensembles.
Most of the ensembles have been generated along lines of constant sum of the three
bare quark masses tuned such that they go through the point defined by the
physical values of the masses of the pion and kaon as well as the flow scale $t_0$. ~\cite{Bruno:2014jqa,Mohler:2017wnb}. These are supplemented by ensembles along lines of constant strange-quark mass~\cite{Bali:2016umi}.
On most ensembles, open boundary conditions in time are imposed to avoid the freezing
of the topological charge as the continuum is approached~\cite{Luscher:2010we,Luscher:2011kk}.

The {\tt openQCD} code is employed. It uses Hasenbusch's frequency splitting for the simulation 
of the light quarks~\cite{Hasenbusch:2001ne,Hasenbusch:2002ai}, 
factorizing \Eq{eq:tm} according to
\be
\begin{split}
  &\det\left\{\frac{[ \hat D^\dagger(m)\hat D(m)+\mu_0^2 ]^2 }{\hat D^\dagger(m)\hat D(m)+2\mu_0^2}\right\}\\
  &=\det\left\{\frac{ \hat D^\dagger(m)\hat D(m)+\mu_0^2 }{\hat D^\dagger(m)\hat D(m)+2\mu_{0}^2}\right\}
 \prod_{n=0}^{N_\mathrm{hb-1}} 
\det\left\{\frac{ \hat D^\dagger(m)\hat D(m)+\mu_n^2 }{\hat D^\dagger(m)\hat
  D(m)+\mu_{n+1}^2}\right\}\\
&\times \det \{\hat D^\dagger(m)\hat D(m)+\mu_{N_\mathrm{hb}}^2 \}\,,
\label{eq:hb}
\end{split}
\ee
where $\mu_i<\mu_{i+1}$ are free parameters for $i>0$. Each factor is introduced by pseudofermions
in the standard fashion. The terms with the smaller $\mu_i$ will be dominated by the contribution of the part of the spectrum
of $D(m)$ with smaller eigenvalues, with the last factor containing the contribution of the UV modes.

Also the RHMC is implemented with frequency splitting in mind. The rational
approximation $R(x)$ is decomposed into factors 
\be
R(\hat Q^2)=A\prod_{k=1}^{n_\mathrm{rat}}  \frac{\hat Q^2+\bar\nu_k^2}{\hat Q^2+\bar\mu_k^2}
=A \frac{\hat Q^2+\bar\nu_1^2}{\hat Q^2+\bar\mu_1^2}\times\prod_{k=2}^{n_\mathrm{rat}} \frac{\hat Q^2+\bar\nu_k^2}{\hat Q^2+\bar\mu_k^2}\,,
\label{eq:rhmc}
\ee
with $\hat Q=\gamma_5 \hat D$.
This is the decomposition in which the pole corresponding to the smallest
$\bar \mu_i$ has been separated, but factorizations into a larger number 
of terms have been used. Such a factorization makes it possible to integrate 
these terms  on a coarser time scale, a fact that will become important in the further discussion.

\subsection{Molecular dynamics integration}

All simulations discussed in this paper have been performed with a three level
hierarchical integration scheme. The outermost level being a second order
Omelyan integrator~\cite{Omelyan2003272} with $\lambda=1/6$, {\tt OMF2} in the
terminology of {\tt openQCD}. On this level some small fermionic forces are
integrated, most of the time the first one or two terms in the product
\Eq{eq:hb} as well as a few terms corresponding to the smallest values of $\bar
\mu_i$ in the RHMC. For each of its steps, the second level is a fourth order
Omelyan integrator, {\tt OMF4} Eq.~(62) and (71) of \Ref{Omelyan2003272}, on
which the rest of the fermion forces reside. On the innermost level, only the
gauge force is integrated, again with one step of the fourth order integrator
per outer step.

With this setup, one is left to tune the number of outer steps, and which of
the fermionic forces are put on the outermost level. Furthermore $\mu_0$ and
the parameters of the rational approximation need to be
  chosen. For some of the ensembles, these choices vary from run to run. This
is why we distinguish ensembles with otherwise the same physical
  parameters by run ids like {\tt r001}. The outer step
size has been chosen such that the acceptance rate for most runs is above $90\%$.

\begin{table}[tbp]
\begin{center}
\small
\begin{tabular}{ccccccccc}
\toprule
  id &   $\beta$ &  $N_\mathrm{s}$  &  $N_\mathrm{t}$  &  $m_\pi$[MeV] &   $m_K$[MeV] &  $m_\pi L$ & bc & $\nneg\neq0$\\
 \midrule
  U103 & 3.40 & 24 & 128	   	& 420 &420  & 4.4 & obc&y\\
  H101 & & 32 & 96	   	& 420 &420  & 5.9 & obc & n\\
  U102 & & 24& 128  	 	& 350 &440  & 3.6& obc&y\\
  H102 & & 32 & 96  	 	& 350 &440  & 4.9& obc&n\\
  U101 & & 24 & 128	 	& 280 &460  & 3.0& obc&n\\
  H105 & & 32 & 96	   	& 280 &460  & 3.9& obc&y\\
  N101 & & 48 & 128    & 280& 460 & 5.9& obc&y\\
  C101 & & 48 & 96	  	& 220 &470  & 4.6& obc&y \\
  D101 & & 64 & 128             & 220 &470  & 6.1& obc&y \\
  H107 & & 32 & 96              & 370 &550  & 5.1& obc&n \\   
\midrule		                        
  B450    &  3.46 & 32 & 64 & 420 & 420&5.2& pbc&n\\
  S400    &       & 32 & 128 & 350 & 440&4.3& obc&y\\
  N401    &       & 48 & 128& 290 & 460&5.3& obc&y\\
  B451    &       & 32 & 64 & 420 & 570 & 5.2 & pbc & n\\ 
  B452    &       & 32 & 64 & 350 & 550 & 4.3 & pbc & n\\
\midrule                            
  H200 & 3.55 & 32 & 96    & 420 &  430 & 4.4 & obc & n \\
  N202 &      & 48 & 128	 & 410 &  410 & 6.4 & obc & n	\\
  N203 &      & 48 & 128	 & 350 &  440 & 5.4 & obc & n	\\
  S201 &      & 32 & 128         & 280 &  460 & 2.9 & obc & n   \\
  N200 &      & 48 & 128	 & 280 &  460 & 4.4 & obc & n	\\
  D200 &      & 64 & 128	 & 200 &  480 & 4.2	& obc	& n \\
  E250 &      & 96 & 192         & 130 &  490 & 4.1   & pbc & y$^*$ \\
\midrule		                        
  N300 & 3.70 & 48 & 128	 & 420	&420  & 5.1 & obc & n	\\
  N302 &      & 48 & 128	 & 350  &460  & 4.2 & obc	& n \\
  J303 &      & 64 & 192	 & 260	&470  & 4.2	& obc & y \\
  E300 &      & 96 & 192         & 180  & 490  & 4.3 & obc & n$^*$ \\
\bottomrule
\end{tabular}
\caption{\label{tab:ens}Subset of CLS ensembles
  which have been investigated for negative real modes. In the id, the letter
  gives the geometry, the first digit the coupling and the final two label the quark mass combination. The lattice spacing as 
  determined in \Ref{Bruno:2016plf} is $a\approx 0.086\,\fm$, $0.0076\,\fm$,
  $0.064\,\fm$ and $0.05\,\fm$ for $\beta=3.4$, $3.55$, $3.46$ and $3.7$,
  respectively. The column marked ``bc'' specifies the boundary conditions in
  time, which can be either open (obc) or periodic (pbc). Ensembles marked with an asterisk in the last column are
    still in production. For more precise estimates of the pion and kaon masses
    including uncertainties please refer to other CLS publications cited in the text. } 
\end{center} 
\end{table}

\section{Determination of the negative real modes\label{sec:neg}}
The obvious method to determine whether the Dirac operator on a given gauge field configuration
has negative real eigenvalues would be to compute the smallest eigenvalues of this operator.
Unfortunately, the methods for such complex systems are quite inefficient and we therefore
resort to studying the spectrum of the Hermitian system $\hat Q = \gamma_5 \hat D$. 
For the numerical examples we use the {\tt PRIMME} package~\cite{PRIMME,svds_software}.

While there is no one-to-one correspondence between the spectra of the two operators, we immediately
note that zero~modes of $\hat D$ are also zero~modes of $\hat Q$. Since
$D(m)=D(0)+m$, for increasing  quark mass the negative
real eigenvalues will first decrease in magnitude before going through zero. 
At this point also a single eigenvalue of $\hat Q$ will change sign. Note that since $\det D_\mathrm{oo}(x)>0$ always,
increasing the mass will not make it zero.

For sufficiently large quark mass, $\hat Q$ has an equal
number of positive and negative eigenvalues. Real negative eigenvalues of $\hat D$ therefore manifest themselves as an asymmetry 
in the number of positive versus the number of negative eigenvalues of $\hat Q$.

Unfortunately, the spectral asymmetry is too difficult to determine directly. 
We therefore use a combination of two indicators.
Firstly, according to the Feynman-Hellmann theorem, for a given eigenvector $\psi$ of $\hat Q$ with eigenvalue $\lambda$
\be
\frac{d}{dm}\lambda = (\psi, \gamma_5 \psi)\,.
\label{eq:dm}
\ee
If this derivative is positive (negative) for positive (negative) eigenvalues, they are moving
away from zero. They are therefore unlikely to go through zero for a larger step in $m$.
Of course, this diagnostic can be misleading because of mixing between the modes of $\hat Q$.

\subsection{Identification of modes as function of mass}
A more reliable way of identifying eigenvalues which cross zero with
increasing mass exploits that the eigenvectors of $\hat Q$ themselves are smooth functions of the mass.
Since the eigenvectors are orthogonal for each value of $m$,  the scalar products
$(\psi'_i,\psi_j)$ will have a modulus close to unity for a matching mode. Here $\psi$ are the normalized
eigenvectors at mass $m$ and $\psi'$ those at $m'$, with $|m'-m|$ reasonably small.

\begin{figure}
  \includegraphics[width=\textwidth]{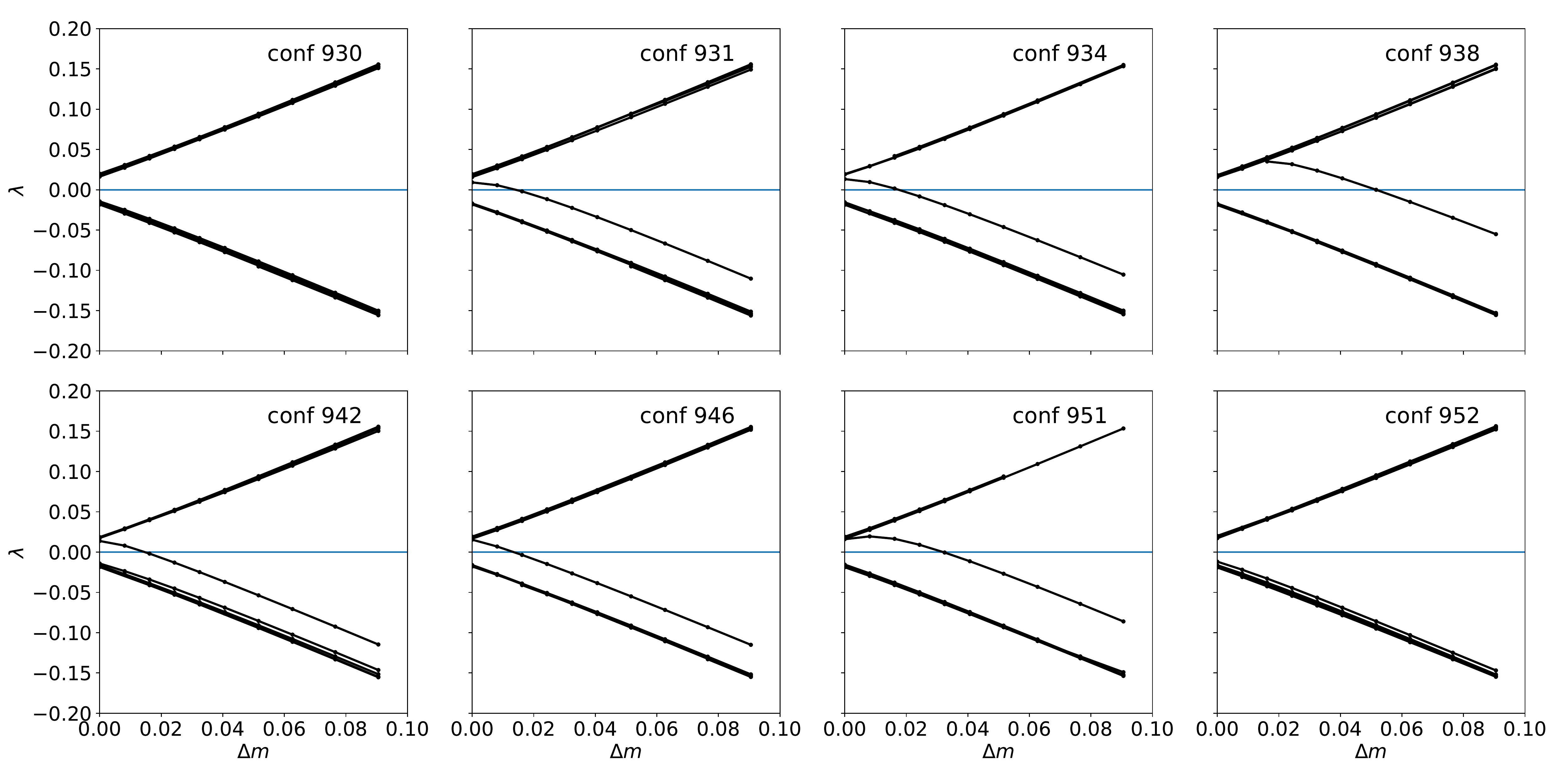}
\caption{Examples of the identification of eigenvalues of $\hat Q$ crossing as
  the strange-quark mass is increased by $\Delta m$.
  We plot measurements on close-by configurations from the {\tt S400r001} run. The value of $\Delta m$ at which zero is crossed
  is a measure for the eigenvalue $-\lambda$ of $D(m_\mathrm{s})$. Note that crossings occur when the other eigenvalues have at least doubled
  their value, again indicated by the increased gap in the other eigenvalues.  \label{f:track}}
\end{figure}

It can also be advisable to combine the two techniques: using a Newton iteration based on 
\Eq{eq:dm}, the zero crossing can be efficiently and unambiguously found. However, even without
this explicit confirmation, the slow variation of the eigenvectors turned out to be a robust tool.
Examples of the tracking of these modes can be seen in \Fig{f:track}. We display a sequence of configurations from 
run {\tt S400r001}, which is in general quite typical, but also
contains a rather extreme excursion to very large negative real value. 

The figure shows the $10$ eigenvalues of $\hat Q$ closest to the origin as a function of the quark mass. The zero point of the x-axis corresponds to the
strange-quark mass and the mass increases towards the right by $\Delta m$. The eigenvalues which are identified between 
successive masses are connected by straight lines, a line crossing zero at $\Delta m_1$ corresponds to a negative real eigenvalue
$-\Delta m_1$ of $D(m_\mathrm{s})$.
The configurations in the run are two trajectories of length $\tau=2$ molecular dynamics units apart. 
First of all, we observe that the transition from positive to negative eigenvalue occurs rather sudden. The intersection
point at the first configuration after the jump is not particularly small,
occurring at a point where the spectral gap is more than 
double its value at the strange-quark mass.

Furthermore, the value of the negative real eigenvalue is moving quite quickly, with a rather large excursion
to more negative values.
This illustrates that one needs to go to relatively large mass shifts for a reliable determination of $\nneg$.

\subsection{Numerical results}

The result of the analysis for ensembles with $\nneg>0$ is collected in \Tab{t:neig}. We give the
probability to find a configuration with a negative strange fermion determinant
$\langle \nneg \rangle$. We note that we did not encounter a single
configuration with two negative real eigenvalues of the strange Dirac operator.
In general, we find that the problem occurs so rarely that it is difficult 
to quantify the probability of configurations with negative real eigenvalues.
Taking into account all configurations investigated. and without proper error analysis, we observe $\langle \nneg \rangle$ of roughly two percent
at $\beta=3.4$ and $\beta=3.46$, on around $0.3\%$ of the configurations at $\beta=3.55$
and on only very few of the configurations ($\sim0.05\%$) at $\beta=3.7$.
As one would expect, negative real eigenvalues quickly become unlikely as the continuum 
limit is approached.

Of course, such a global view ignores the varying physical parameters, in particular the
quark masses as well as the details of the 
algorithmic choices, like the twisted-mass reweighting parameter $\mu_0$, and 
the rational function used in the RHMC. We will discuss their impact in the next section.

Furthermore, the above statement ignores the impact of autocorrelations on such numbers.
As becomes evident from \Tab{t:neig}, large values for $\langle \nneg \rangle$ come with
large autocorrelations in this quantity: once a negative real eigenvalue occurs, the 
Markov chain frequently gets stuck and therefore many such configurations are produced in a row.
This does not mean, that these configurations are particularly likely. Indeed, the uncertainties
of $\langle \nneg \rangle$ are typically on the order of $100$\%. It simply means that
the runs are too short to determine these values precisely. 

\begin{figure}
  \includegraphics[width=\textwidth]{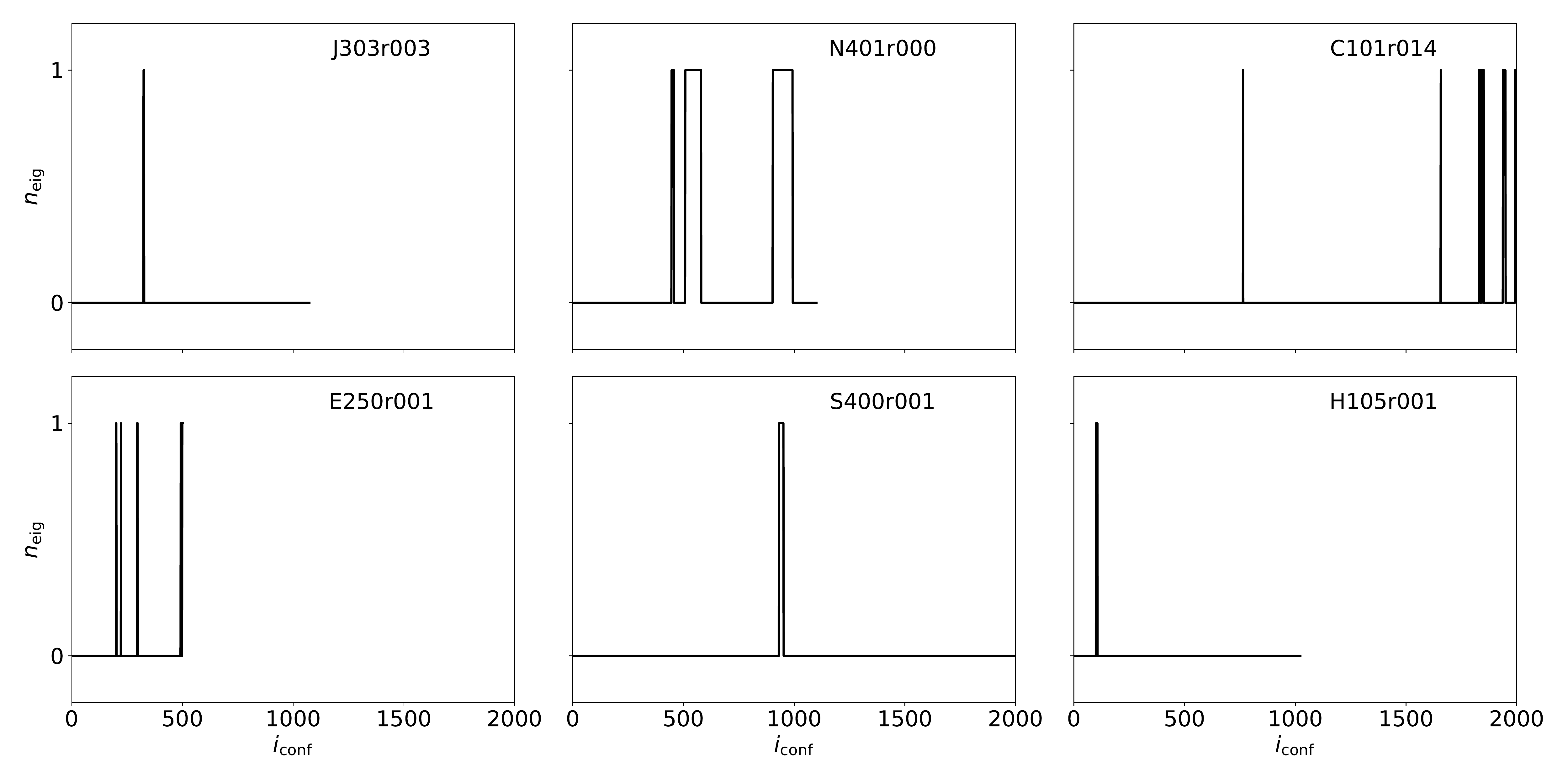}
  \caption{\label{f:mc}Monte Carlo time history of the number of negative real eigenvalues of the 
  Dirac operator. Configurations are spaced by 4 units of molecular dynamics time; only for J303 this
  spacing amounts to 8 units. We observe that changes in the number are rare.}
\end{figure}

To illustrate this issue, we have collected Monte Carlo time histories of the number of negative eigenvalues
in \Fig{f:mc}. In fact, we find only few ensembles, where $\nneg>0$ has been visited more than once.

\begin{table}[tbp]
  \begin{center}
  \begin{tabular}{lllllllllll}
    ens & $\langle \nneg \rangle$ & $\tau_\mathrm{int}$&$n_\mathrm{sec}$&$n_\mathrm{max}$& stat& $\mu_0$ & $n_\mathrm{out}$ & $n_\mathrm{rat}$ & $r_a, r_b$   \\
    \hline
    H105r001& 0.004(4)& 2.0(4) &2 &2 &1023 & 0.001 & 10 & 11 & [0.01,7.3] \\
    H105r002& 0.001(1)&0.50(3)&1&1&1042&  0.001 & 10 & 11 & [0.01,7.3]\\
    H105r005& 0.11(11)&45(22)&1 &92&837 & 0.0005 & 7 & 13 & [0.0032,7.6]\\
    N101r001& 0.11(11)& 15(7)&1 & 30 &280 & 0.0005&9&14&[0.0032,7.6]\\
    C101r014& 0.02(1) & 6(1)   &8  &12 &2000& 0.0006 &12&13&[0.006,7.8]\\
    C101r015& 0.09(9) & 25(12) & 6 & 33&601&0.0003&13&13&[0.006,7.8]\\
    D101r005& 0.03(3) &4(2)&1&9&286&0.0003&12&14&[0.006,7.8]\\
    U103r002& 0.09(9) &76(36)&1&154&1781&0.001&6&12&[0.0056,7.5] \\
    U103r003& 0.0005(5) &0.50(2)&1&1&1819&0.001&6&12&[0.0056,7.5]\\
    U102r002& 0.004(4) & 7(1) & 2 & 12 & 3562 & 0.002 & 6 & 12 & [0.007,8.0]\\
    \hline
    N401r000& 0.16(9)  &35(16)& 3&90&1100 & 0.00065 & 8 & 14 & [0.002,7.5]\\
    S400r000& 0.01(1)  &5(1) &1&10&872& 0.00065 & 7 & 12 & [0.01,7.3]\\
    S400r001& 0.01(1)  &11(2)&1&21&2001& 0.00065 & 7 & 12 & [0.01,7.3]\\
    \hline
    E250r000& 0.12(12) &8(4) & 4 & 16 & 151 & 0.0001 & 14 & 14 & [0.01,7.5]\\
    E250r001& 0.03(2) & 3(1) & 4 & 5 & 503 & 0.0001&15&14&[0.01,7.5]\\
    \hline
    J303r003& 0.003(3) &1.5(3)&1&3&1073&0.00075&6&13&[0.008,7.0]&\\
\end{tabular}
  \end{center}
  \caption{Runs in which at least one config with $\nneg\neq 0$ occurred. The probability of this event is given, 
  together with its autocorrelation time. Note that the errors of these numbers have a very large uncertainty due to the 
  long autocorrelations observed in some of the runs. As another measure of
  autocorrelations we give the number of sections in the chain with $\nneg\neq
  0$ denoted by $n_{sec}$,
  the length of the longest such region $n_{max}$ and the total statistics in number of configurations. 
  We also list the twisted mass parameter $\mu_0$, the number of outer
  integrator steps $n_{out}$, and the parameters of the rational
    approximation. We observe a correlation of a lower number of
    outer integrator steps and longer stretches of $\nneg\neq 0$.
  \label{t:neig}}
\end{table}

\section{Discussion}
In the previous section, we have established that, in particular on the 
coarser lattices, there is a non-negligible frequency with which negative
real eigenvalues of the strange Dirac operator occur.

A possible explanation of this occurrence could be the modification of the 
action by twisted-mass reweighting and the chosen RHMC function. Both
increase the probability for small eigenvalues to appear in the ensembles 
(before applying the reweighting). However, we find that the mass shifts at 
which the crossings occur are typically large, of the order of twice the
the spectral gap $\langle |\lambda_\mathrm{min}|\rangle$. This means that the negative
eigenvalues of $D$ typically have (at least) the same magnitude as the smallest positive ones.
Most of them are in a region, where the approximated action and the ``exact'' 
fermion determinant are almost the same. For the RHMC, the reweighting factors
typically fluctuate on the percent level and, also for the twisted-mass reweighting
factors, we do not observe a correlation between small values and
a non-vanishing $\nneg$.

Another possibility would be algorithmic instabilities or even an error in the 
simulation code. While both are hard to exclude, the rather consistent 
picture with the number of negative eigenvalues 
decreasing rapidly towards the continuum
and no apparent correlation with volume, integrator step size, or acceptance
rate does not  make this a very convincing explanation.

Therefore our current working hypothesis is that the phenomenon described in this paper
is a feature of the action. The fluctuations in the spectrum are simply larger 
than naively expected. The setup with the regularized actions is therefore necessary
to reach these areas of configuration space and is essential 
for a correct simulation.

\subsection{Autocorrelations}
As already mentioned above, even with the regularized action a negative real 
eigenvalue has to cross a barrier in the action around $\lambda=0$ in order to
disappear. This is linked with larger, fluctuating forces.
In how far this is a problem depends on the choice of the integrator for the 
molecular dynamics equation of motion. With a precise enough
  integrator for the terms which are dominated by the contribution 
  from small eigenvalues,  the right momentum to get across the barrier is still
  needed, but the acceptance of such a trajectory will not suffer.

In the practical simulations on which this analysis is based, the force
terms which receive most contributions from the smallest eigenvalues 
are most of the time tiny. They have therefore been integrated on a 
coarser time scale. If the associated step size if too large this can lead
to poor acceptance of trajectories in which an eigenvalue has become very small or
crossed zero.

\begin{figure}
  \begin{center}
\includegraphics[width=0.45\textwidth]{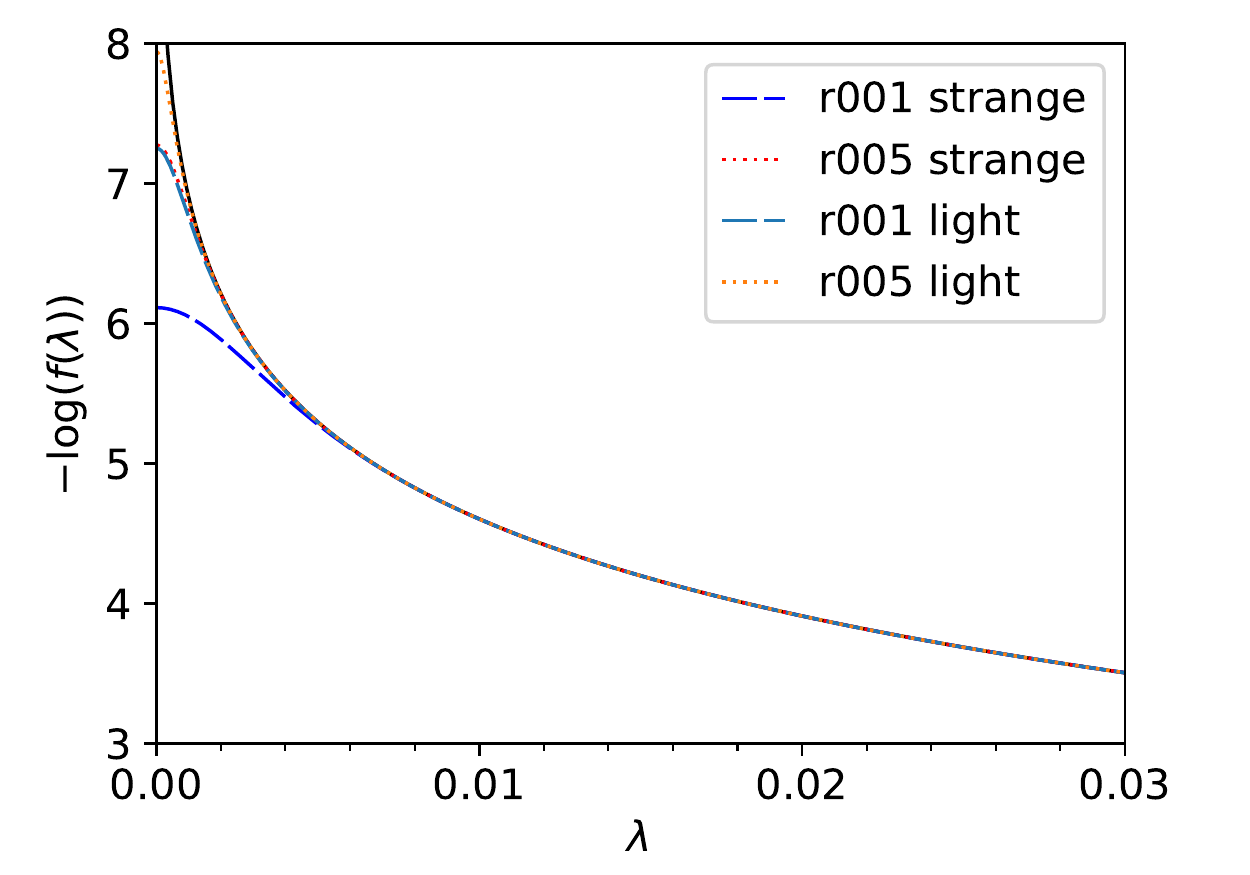}
\includegraphics[width=0.45\textwidth]{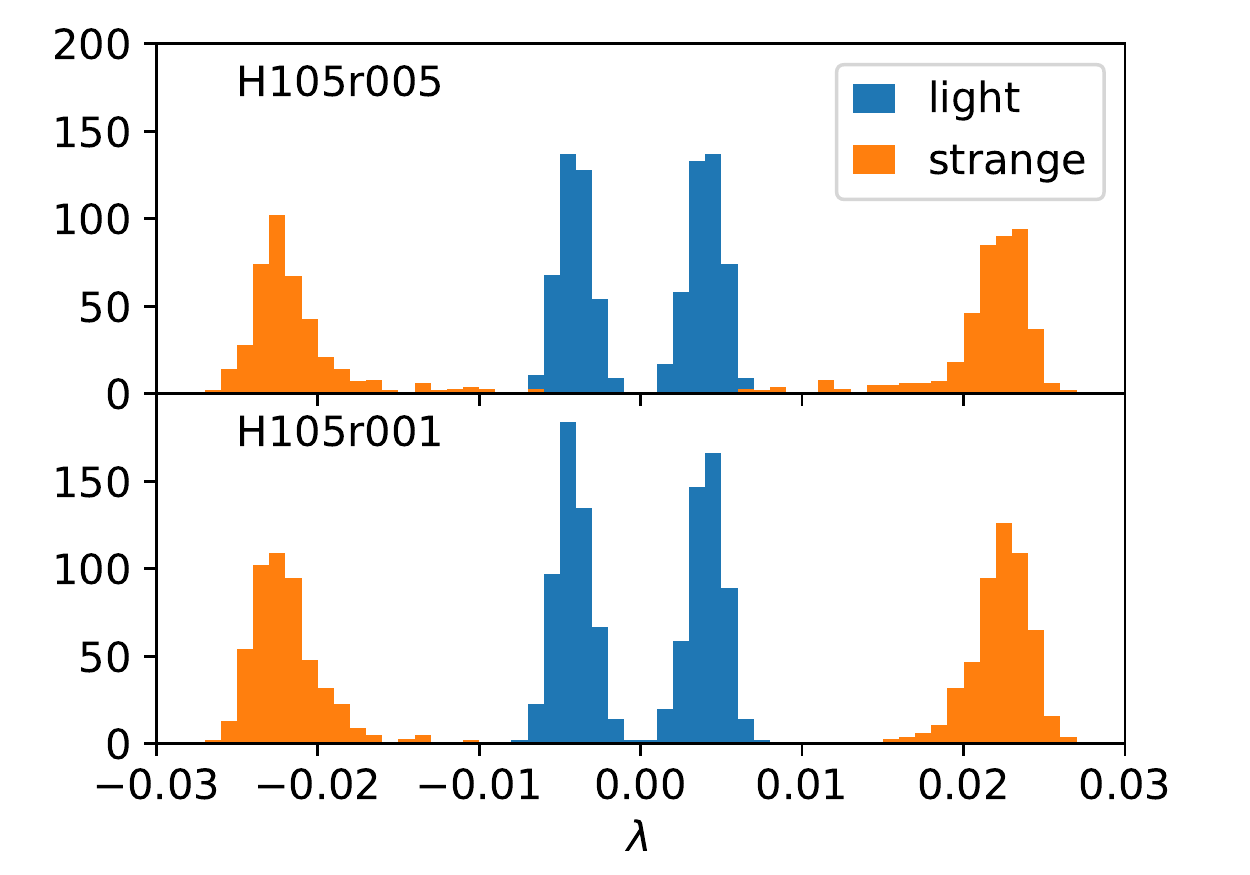}
  \end{center}
  \caption{On the left, function for a single variable which replaces the $-\log(x)$ of 
  the fermion action, for the light quark due to the twisted mass reweighting and for the 
  strange due to the RHMC. We show the parameters of the {\tt H105r001} and {\tt H105r005} runs. Note that 
  for the latter the potential barrier is roughly 1 unit higher.
  On the right we give the measured distribution of the lowest eigenvalue of $\hat Q$, not
  including the reweighting factors and without giving uncertainties, which are difficult to compute in the case
  of the {\tt H105r005} runs due to the large autocorrelations discussed in the text.
\label{f:act}}
\end{figure}

That the tuning of such a setup can be delicate
when there is a nonvanishing probability of
negative real eigenvalues, is illustrated with two runs from the H105
ensemble.
In runs {\tt r000} and {\tt r001}, $\mu_0=0.001$ has been chosen along with an 11 pole
rational approximation in the interval $[0.01,7.3]$. With the first term
in the light quark product \Eq{eq:hb} and the two smallest poles of the RHMC on the
outermost level, $10$ steps have led to an acceptance rate of $97\%$. 
This has to be compared to the choices taken for the r005 run. Here
  $\mu=0.0005$ along with a 13 pole rational function in $[0.0032,7.6]$ was used.
Choosing 7 outer integration steps, again the first term of the light fermion forces and the 
three smallest $\bar \mu_i$ in the RHMC on the outer level, has led to an
acceptance rate of $89\%$. We plot the corresponding regularizing functions in \Fig{f:act}.

\begin{figure}
  \begin{center}
\includegraphics[width=\textwidth]{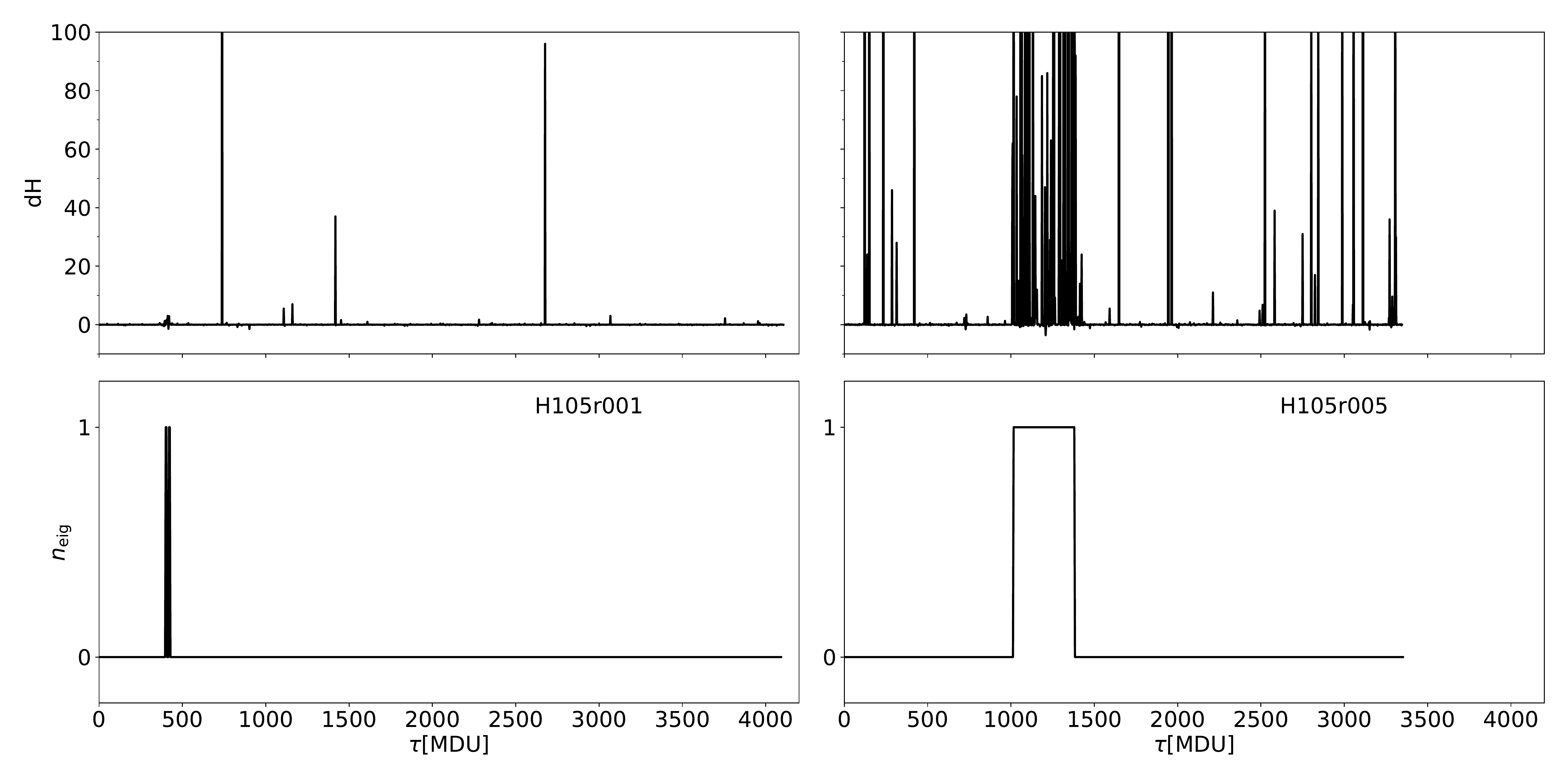}
  \end{center}
  \caption{The violation of the HMC Hamiltonian in the top row is confronted with the 
  number of negative eigenvalues of the Dirac operator in the bottom. We show the data
  for the runs H105r001 and H105r005. There is clear evidence that the period of poor acceptance,
  i.e. small $\exp(-dH)$ is correlated with the appearance of a negative real eigenvalue in the
  latter run. The different choice of molecular dynamics integrator might be responsible for the
  system staying in this state for an extended time.\label{f:dh}}
\end{figure}

At first sight, the three chains are equally acceptable. The reweighting factors were
well behaved, with the twisted mass reweighting factors fluctuating less for r005 a expected.
No signs of strange behavior in elementary observables were detected either.
Only in the history of the energy violation of the molecular dynamics depicted in \Fig{f:dh}, a
long stretch of trajectories between number 1000 and 1500 with 
low acceptance has been observed.

It turns out that in this region, a negative real eigenvalue of the strange Dirac operator
had developed. In this situation, it matters how accurately the terms with the smallest $\hat \mu_i$
are integrated, as they are dominated by the contributions from the smallest eigenvalues.
Here we note that for the r005 run, we put more poles on the outer level of the integrator and 
we increased its step size significantly. The likely explanation is that trajectories
which would have changed this eigenvalue's sign were rejected, such that the system did not get out of 
this region.

Note that also in the other runs negative real eigenvalues were produced, but
the system quickly moved away from these configurations. While these
considerations are based on very low statistics, unavoidable by the nature of
the problem, they do give a consistent picture. Note that from the point of
view of tuning of the algorithm, this is an unpleasant situation, since the
problem can even arise only after a significant portion of the full run. In
hindsight, it would probably be better to use only a two-level scheme and not
try to save computing time by using a further integrator level for these problematic forces.

\subsection{Regularizing the action}
As we have seen, the choice of the rational approximation and the quality of the
integration of the associated forces has a determining impact on the problem
discussed in this paper.
Since one would naively expect that at a quark mass as large as the strange's, the
spectrum of $\hat Q^2$ can be confined to a region $[r_a,r_b]$ in a 
practical simulation. Given the fast convergence of the Zolotarev approximation
to $1/\sqrt{x}$, one then aims at choosing the number of poles and the range
given by $r_a$ and $r_b$ such that no eigenvalue outside this interval
will be encountered and the approximation error is negligible.

As has become clear from the findings presented here, this is a dangerous
choice in that, both, decreasing $r_a$ and increasing $n_\mathrm{rat}$ improve
the approximation but also heighten the obstacle at $\lambda=0$, which the eigenvalues
have to overcome in order to get out of a region with a negative real eigenvalue.

These considerations can help understand the different behavior between the runs {\tt H105r001} and {\tt H105r005}
discussed in the previous section. In \Fig{f:act}, we show as an illustration
the function by which $-\log(\lambda)$, the contribution to the action of a single eigenvalue, is replaced due to the 
twisted mass regularization and the rational approximation. This needs to be compared 
with the distribution of the lowest eigenvalue given in the same figure.

As we see, for the strange quark the approximations do not differ in the region
of the lowest eigenvalues of $\hat Q$. However, the maximum of the depicted 
functions at $\lambda=0$ is about one unit higher, making in much less likely
to be overcome in a trajectory. Given these considerations, it seems advisable
to fix $r_a$ and $r_b$ such that it covers the spectrum of $\hat Q^2$ apart from 
maybe some rare outliers which seem unavoidable in light of our findings while using 
a low order rational approximation such that one gets
  an acceptable fluctuation in the associated reweighting factors.

\subsection{Light quarks\label{s:light}}
Negative real eigenvalues of the Dirac operator at the strange mass are also
negative at the light-quark mass (if chosen below the strange's mass-value).
While the degeneracy of the two quark masses makes the product of the two
determinants positive, the negative eigenvalues can still introduce large
autocorrelations if the algorithm is inefficient in changing their sign.

In \Fig{f:act} we compare functions with which the $-\log(\lambda)$ has been replaced
in our setup for the runs {\tt H105r001/2} and {\tt H105r005}. 
As can be seen, the fixed rational approximation acts quite similarly to the twisted mass reweighting,
by chance the twisted mass parameters on the former runs give a function for one flavor which is almost identical 
to the rational approximation chosen for the latter.

Of course, it is not possible to deduce from the function alone whether or not there will be autocorrelation
problems. It also depends on the typical distribution of the
eigenvalues among other factors.
In our analysis of the two algorithmic setups, we observe negative real eigenvalues
of the light Dirac operator also on configurations, where this is not the case 
at the strange mass. However, also at the light-quark mass, at most one negative
real eigenvalue occurs. These negative eigenvalues at the light-quark mass do not
exhibit significant autocorrelations and (with the very limited statistics) on
{\tt H105r001} are roughly twice as likely as at the strange-quark mass.

\subsection{Effect of reweighting}
With the (additional) reweighting factors equal to $\pm 1$ and the rare occurrence of the 
$-1$, it is obvious that their effect on the final observables will be small, at least as
long as there is no strong correlation between the observable an the reweighting factor.
If there is no correlation between the observable and reweighting factor, then the variance
is not affected at all, see Eq.~(5.4) of \Ref{Bruno:2014lra}. However, the autocorrelation 
problem discussed above can significantly impact the achievable accuracies, if a reliable
error analysis is possible at all.

To illustrate the effect in a real situation, we again took the ensembles {\tt H105r001} and {\tt H105r002}
and compared them to {\tt  H105r005}. These have the same physical parameters, but in the latter run the 
problem of negative real eigenvalues is much more pronounced. 

For the pion mass, we extract $am_\pi=0.1213(12)$ from the former two ensembles
and $am_\pi=0.1207(21)$ from the latter when we take the effect of the
determinant's sign into account. This is to be confronted with $0.1213(12)$ and
$0.1220(17)$, respectively, without this reweighting factor. As expected, for
the ensembles with very few negative signs of the determinant there is no
significant difference, neither for the value, nor
for the uncertainty. For {\tt r005}, we observe a shift in the
value which is somewhat smaller than the
  statistical uncertainty and also an increase in the error.

For the pseudoscalar decay constant the situation is essentially the same:
on the first two ensembles it changes from 0.0763(10) without the signs to
0.0764(10) including them. On ${\tt r005}$ the respective numbers are 0.0758(9)
and 0.0748(12).

\section{Conclusions}

On coarser lattices, the non-perturbatively improved Wilson Dirac operator at
the strange-quark mass features negative real eigenvalues on a non-negligible
subset of the configurations in the 2+1 flavor CLS ensembles. This has not been
anticipated during the planning of the simulations, but the corresponding sign
can be included in the measurement as a reweighting factor. We have described a
robust but expensive way to compute this sign, however, it is difficult to
exclude that occasionally a negative real eigenmode of
the Dirac operator remains undetected. Since the
effect of the reweighting seems to be rather small, we however assume that the
effect of potentially missing a few such configurations would be even smaller.

The scenario observed also has consequences on the planning of the simulations.
Specifically one needs to ensure that all regions of configuration space
can be reached by the algorithm, even if one has the prejudice that some  regions 
are not ``relevant''. In the CLS ensembles, twisted-mass reweighting for the 
light quarks and the RHMC with fixed rational functions are employed to this
end. However, for some of the runs, in hindsight, one should have used a larger 
value for the twisted mass parameter and/or fewer poles and smaller approximation 
range for the rational function. This would have made the transition of eigenvalues
through zero easier and reduced autocorrelations. Also an integration scheme
for the molecular dynamics equations of motion, where the forces dominated by 
  the contributions of small eigenvalues of the Dirac operator are not put on
  a very coarse step size, seems advisable.

Note that one part of our action
is not regulated: the diagonal term of the Dirac operator $\det D_\mathrm{oo}$.
The fact that it turns out to be always positive might be due to the infinite barrier
at zero determinant, rather than the actual physics of the system. This is impossible
to tell after the simulation. As a side remark: this determinant is constant for 
unimproved Wilson fermions or the variant proposed in \Ref{Francis:2019muy}.

The practical ergodicity of Monte Carlo simulations remains difficult to assess
in general, and it will always depend on the discretization and algorithms in question
where possible difficulties might arise.  Our discussion also
highlights the fact that typical lattice simulations are not close to the continuum
if it comes to details of the behavior of single eigenvalues of the Dirac operator.

\acknowledgement
We thank our CLS colleagues for providing the gauge field configurations
this study is based on. We are indebted to Gunnar Bali, Mattia Bruno, Marco C\`e, Georg von Hippel,
Martin L\"uscher, Teseo San Jos\'e, Rainer Sommer, and Hartmut Wittig for very useful
discussions. Eigenvalue calculations were performed on the HPC clusters
``HIMster II'' at the Helmholtz-Institut Mainz and ``Mogon II'' at JGU Mainz.

\usebiblio{latt,ALPHA}

\end{document}